\documentstyle[aps,twocolumn,epsfig,eqsecnum]{revtex}

\begin{document}

\newcommand{\de}{^{o}} \newcommand{\eq}{equation}
\newcommand{\beq}{\begin{equation}} \newcommand{\eeq}{\end{equation}}
\newcommand{\bea}{\begin{eqnarray}} \newcommand{\eea}{\end{eqnarray}}
\newcommand{\pri}{^{\prime}} \def\alt{\,\raise 0.6ex\hbox{$<$}\kern
-0.75em\lower 0.47ex
   \hbox{$\sim$}\,} \def\agt{\,\raise 0.6ex\hbox{$>$}\kern
-0.75em\lower 0.47ex
    \hbox{$\sim$}\,} \newcommand{\real}{{\cal R}}
\newcommand{\bs}{{\vec{S}}} \newcommand{\bls}{{\bf{s}}}
\newcommand{\bst}{{\bf{S}}_{tot}}

\newcommand{\dj}{\delta J} \def\jour#1#2#3#4{{#1} {\bf #2}, #3 (#4).}
\def\tit#1#2#3#4#5{{#1} {\bf #2}, #3 (#4).} \def\epl{Europhys. Lett.}
\def\prl{Phys. Rev. Lett.} \def\pr{Phys. Rev.} \def\prb{Phys. Rev. B}
\def\jpco{J. Phys. Cond. Mat} \def\jpc{J. Phys. C} \def\jap{J. Appl.
Phys.} \def\zpb{Z. Phys. B}

\newcommand{\mubf}{\mbox{\boldmath $\mu$}}

%\setlength{\baselineskip}{0.5cm}

%\begin{titlepage}
%\twocolumn[\hsize\textwidth\columnwidth\hsize\csname @twocolumnfalse
%\endcsname

\twocolumn[\hsize\textwidth\columnwidth\hsize\csname    %<--------
@twocolumnfalse\endcsname                               %<--------

\begin{title} {\Large \bf Origin of Spin Ice Behavior in Ising
Pyrochlore Magnets with Long Range Dipole Interactions: an Insight from
Mean-Field Theory} \end{title}

\author{Michel J.P. Gingras$^{1,2}$ and Byron C. den Hertog$^1$}
\address{$^1$Department of Physics, University of Waterloo, Ontario
Canada N2L 3G1} \address{$^2$Canadian Institute for Advanced Research,
180 Dundas Street West,  Toronto, Ontario,  M5G 1Z8, Canada}

\vspace{3mm}

\date{\today} \maketitle

\begin{abstract}
Recent
experiments suggest that the Ising pyrochlore magnets
${\rm Ho_{2}Ti_{2}O_{7}}$ and
${\rm Dy_{2}Ti_{2}O_{7}}$ display qualitative
properties of the ferromagnetic
nearest neighbor
spin ice model
proposed by Harris {\it et al.},
Phys. Rev. Lett.  {\bf 79}, 2554 (1997).
The manifestation of
spin ice behavior in these systems
{\it despite} the energetic
constraints introduced by the strength and
the long range nature of
dipole-dipole interactions,
remains difficult to understand.
We report here results from
a mean field analysis that
shed some light on the origin of spin ice behavior in $(111)$ Ising
pyrochlores.
Specifically, we find that there exist a large frustrating effect
of the dipolar interactions
beyond the nearest
neighbor, and that the degeneracy established by effective ferromagnetic
nearest neighbor interactions is only very weakly lifted by the long
range interactions.
Such behavior only appears beyond a cut-off distance corresponding to
$O(10^2)$ nearest neighbor.
Our mean field analysis shows that truncation of dipolar interactions
leads to spurious ordering phenomena that change with the
truncation cut-off
distance.\\

To appear in Canadian Journal of Physics for the Proceedings of the
{\it Highly Frustrated Magnetism 2000 Conference}, Waterloo, Ontario,
Canada, June 11-15, 2000
\\

\end{abstract}

\vskip2pc]                                     %<---------

\vspace{4cm}

%\end{titlepage}

%\newpage

%\pacs{PACS numbers: 75.10.Hk, 75.40.Mg, 75.40.Gb} %]

%\input{psfig}

\section{Introduction}

The past five years have seen a resurgence of significant interest
devoted to the systematic study of geometrically frustrated
magnetic systems \cite{review-fafs,chandra,shender_rev,henley_rev,moessner_rev}.
Frustration arises when a
magnetic system cannot minimize its total classical ground-state energy
by minimizing the energy of each spin-spin interaction
individually~\cite{toulouse}. This most often occurs
in materials containing
antiferromagnetically coupled magnetic moments that reside on
geometrical units, such as triangles and tetrahedra, which inhibit the
formation of a collinear magnetically-ordered state.

It is very common for models of highly frustrated magnetic systems
to display ground state
degeneracies where the system is ``underconstrained'', giving
many distinct spin configurations give the same ground state
energy. The best known example is probably that of the
two-dimensional Ising antiferromagnet on a simple triangular
lattice, for which it was shown by Wannier that
the ground state is macroscopically
degenerate with an extensive ground state entropy \cite{wannier}.
Another example is the nearest-neighbor antiferromagnet
face-centered cubic lattice
with Heisenberg \cite{FCC0,FCC1,FCC2,FCC3,FCC4}, and
Ising spins \cite{wengel}.
In the past ten years,
much attention has been devoted to the kagom\'e
lattice of corner-sharing triangles, and the pyrochlore lattice
of corner-sharing tetrahedra
 with Heisenberg spins interacting via
nearest neigbor antiferromagnetic
interactions
\cite{review-fafs,chandra,shender_rev,henley_rev,moessner_rev}.

Typically, there are two generic class of mechanisms that can lift
the degeneracies in frustrated magnetic systems. The most common
and simplest mechanism proceeds via
energetic perturbations beyond the term(s) 
in the spin Hamiltonian 
that cause the degeneracies. Degeneracies can also be lifted by
either thermal fluctuations or quantum zero-point fluctuations
\cite{shender_rev,henley_rev,shender,henley_prl,reimers-mc,plumer,bramginreim,random_obd}.
There are a number of well known situations
where spin interactions beyond nearest neighbor cause long range order that
would otherwise be absent in a highly frustrated lattice with
only nearest neighbor antiferromagnetic exchange present.
For example,
the classical kagom\'e~\cite{harris} and
pyrochlore~\cite{reimers-mc,plumer,reimers-mft}
nearest neighbor Heisenberg antiferromagnet
lattices  display either
N\'eel order or a dramatic reduction of ground state degeneracy
when 
exchange interactions beyond nearest neighbor are
considered. Long range dipole-dipole interactions
in the FCC Heisenberg antiferromagnet~\cite{FCC3,FCC4}
and in the pyrochlore lattice with Heisenberg spins (as pertains to the
Gd$_2$Ti$_2$O$_7$ pyrochlore
material~\cite{gd2ti2o7,palmer,bramwell-gd}),
have also been shown to reduce significantly the degeneracy otherwise
present in the nearest neighbor version of these systems.

While high geometric frustration usually arises in 
antiferromagnetic systems, 
Harris and collaborators recently showed that the
pyrochlore lattice of corner sharing tetrahedra
with Ising spins pointing along a local cubic (111) axis  
(See Fig. 1)
constitutes an interesting and
 unusual example of high geometric frustration when the
nearest neighbor interaction is actually
ferromagnetic~\cite{harris-prl,harris-jpc}.
Harris and collaborators introduced the
concept of {\it spin ice} to emphasize the analogy between
ferromagnetically coupled (111) Ising moments on the pyrochlore lattice
and the problem of proton ordering in common
hexagonal ice, I$_{\rm h}$~\cite{harris-prl,harris-jpc,global,anderson}.
In  the simple model of nearest neighbor ferromagnetic interactions,
the system has the same `ice rules' for the  construction of its ground
state as those for the  ground state  of real  ice \cite{bernal},
hence the name {\it spin ice}.
The resulting
extensive ground state entropy of hexagonal ice 
was first estimated
by Pauling in 1935 \cite{pauling} which, remarkably,
provides a fairly
accurate estimate of the residual low temperature entropy
of real pyrochlore spin ice materials \cite{ramirez-nature}.

%%%%%%%%%%%%%%%%%%%%%%%%%%%%%%%%%%%%%%%%%%%%%%%%%%%%%%%%%%%%%

\vspace{-1.5cm}
\begin{figure}
\begin{center}
\includegraphics[width=9cm]{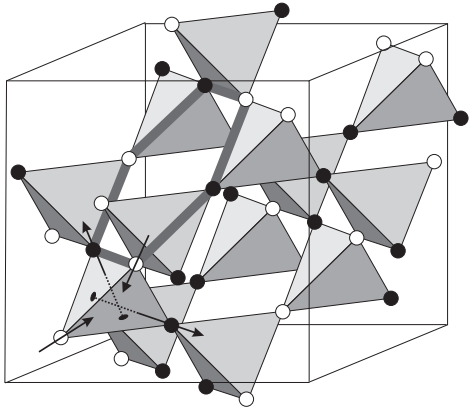}
\vspace{-15mm}
\caption{The lower left `downward'
tetrahedron of the pyrochlore lattice
shows Ising spins (arrows). Each spin axis is along
the local $(111)$ quantization axis, which goes from one site
to the middle of the opposing triangular face (as shown by the disks
on the triangular faces) and meets
with the three other $(111)$ axes in the middle
of the
tetrahedron. For clarity,  black and white circles on the lattice points
denote other spins. White represents a spin pointing into a downward
tetrahedron while black is the opposite. The entire  lattice is shown in
an ice-rules state (two black and two white sites for every tetrahedron).
The hexagon (thick gray line) shows a low energy ``loop'' excitation
(see text)  which corresponds to
reversing all colors (spins) on the loop to produce a new ice-rules state.}
\end{center}
\end{figure}

%%%%%%%%%%%%%%%%%%%%%%%%%%%%%%%%%%%%%%%%%%%%%%%%%%%%%%%%%%%%%

\vspace{-3mm}

Interestingly,
materials with local (111) Ising spins are realized in 
Ho$_2$Ti$_2$O$_7$~\cite{harris-prl,ramirez-prl,ramirez-japp,bramwell-Sq},
Dy$_2$Ti$_2$O$_7$~\cite{ramirez-nature} and Tb$_2$Ti$_2$O$_7$
\cite{tb2ti2o7,gingras-tb2ti2o7,tb2ti2o7-jpsj,tb2ti2o7-except},
where the rare earth magnetic
moments (Ho$^{3+}$, Dy$^{3+}$ and Tb$^{3+}$) reside on the sites of the
pyrochlore lattice.
For $(111)$ Ising pyrochlore systems, each moment points along the axis
joining the centers of the two tetrahedra that it belongs to
(see Fig. 1). For
ferromagnetic nearest neighbor interaction, the ground state is
macroscopically degenerate, but with the constraint that two moments
must point in and two must point out of every tetrahedron which,
as mentioned above, is a constraint
that maps precisely onto the
ice rules~\cite{harris-prl,harris-jpc,global,anderson,bernal,pauling}.

The nearest neighbor model of Harris and
Bramwell \cite{harris-jpc} shows no evidence of a magnetic transition
to long range magnetic order~\cite{global,anderson,champion-prl}, but
displays a broad peak in the specific heat at a temperature of the
order of the nearest neighbor ferromagnetic coupling. Both
Ho$_2$Ti$_2$O$_7$~\cite{harris-prl,ramirez-prl,ramirez-japp,bramwell-Sq} and
Dy$_2$Ti$_2$O$_7$~\cite{ramirez-nature} show qualitative behavior
consistent with this spin ice picture.

As explained above, 
the occurrence of a macroscopic degeneracy in frustrated magnetic
is due to ``underconstraints''; many
different spin configurations (here the ice-rules ``two-in$-$two-out'' per
tetrahedron) give the same ground state energy. We
therefore expect that the microscopic origin
of the spin ice
phenomenon in Ho$_2$Ti$_2$O$_7$ and Dy$_2$Ti$_2$O$_7$ has to directly
involve such
a manifestation of underconstraints in the energetics involved in the
formation 
of the spin ice ground state. 

Our goal is therefore to understand how the competing energy scales
in the spin ice materials are able to produce underconstraints.
 In order to identify
such underconstraints, we must
first consider
the various energy scales at play in the above materials.
Firstly, 
the single-ion ground state for Ho$^{3+}$ and Dy$^{3+}$ in the
pyrochlore structure is well described by an effective
classical Ising doublet with a ground state magnetic moment of $\mu
\approx 10\mu_{\rm B}$ and a nearest neighbor
distance, $r_{nn}$, of  approximately $3.54 \AA$ for both
materials \cite{ramirez-japp}. For two
nearest neighbor moments pointing along their local (111) direction,
the nearest neighbor dipole-dipole energy scale, $D_{nn}$, is
\begin{equation}
\label{eqn1}
D_{nn} =
\frac{5}{3}(\frac{\mu_0}{4\pi})\frac{\mu^2}{r_{nn}^3} \;\; \approx +2.35
{\rm K}
\;\;\; ,
\end{equation}
where,
as discussed in the next section, 
 the 5/3 factor comes from the orientation of the Ising quantization
axes relative to the vector direction connecting
interacting nearest neighbor magnetic moments.
Interestingly, $D_{nn}$ is positive (i.e.
ferromagnetic) and therefore, as in the Harris and Bramwell
nearest neighbor model~\cite{harris-jpc}, is frustrating for Ising spins on a
tetrahedron.

%%%%%%%%%%%%%%%%%%%%%%%%%%%%%%%%%%%%%%%%%%%%%%%%%%%%%%%%%%%%%

The experimentally determined Curie-Weis temperature, $\theta_{\rm CW}$,
 extrapolated
from temperatures below 
$T \sim 100$K is +1.9 K for
Ho$_2$Ti$_2$O$_7$~\cite{harris-prl} and
+0.5 K for
Dy$_2$Ti$_2$O$_7$~\cite{ramirez-nature}, 
respectively. These two values show that
$\theta_{\rm CW}$ is of the same order of magnitude as the nearest neighbor
dipolar energy scale. Furthermore, it is well known that rare-earth
ions possess very small exchange
energies. Consequently, dipole-dipole interactions
in Ho$_2$Ti$_2$O$_7$ and Dy$_2$Ti$_2$O$_7$ constitute a
preponderant interaction,  as opposed to a weak perturbation as in
magnetic transition elements where the exchange interaction predominates
over dipolar interactions.
  As we discuss below,
fits to experimental data show that the nearest neighbor
exchange interaction is antiferromagnetic for both
Ho$_2$Ti$_2$O$_7$ and
for Dy$_2$Ti$_2$O$_7$ and, when considered alone,
is therefore not frustrating for $(111)$ Ising
moments on the pyrochlore
lattice~\cite{bramginreim,harris-jpc,moessner-rapcomm}.
In order to
consider the combined role of exchange and dipole-dipole interactions,
we find it useful to
define an effective nearest neighbor energy scale, $J_{\rm
eff}$, for $(111)$ Ising spins:
\begin{equation}
 J_{\rm eff} \; \equiv \;	 J_{\rm nn}+D_{\rm nn}
\;\;\; ,
\end{equation}
where $J_{\rm nn}$ is the nearest neighbor exchange energy between
$(111)$ Ising moments.
This simple description predicts that a $(111)$ Ising system
would display spin ice properties even for antiferromagnetic
nearest neighbor exchange, $J_{\rm nn}<0$, but has long as
$J_{\rm eff}=J_{\rm nn}+D_{\rm nn}>0$.
Fits to experimental data
give
$J_{\rm nn} \sim -0.52$ K for
Ho$_2$Ti$_2$O$_7$ \cite{bramwell-Sq}
and  $J_{\rm nn} \sim -1.24$ K for Dy$_2$Ti$_2$O$_7$ \cite{denHertog-prl1}.
Thus, $J_{\rm
eff}$  is positive (using $D_{\rm nn}=2.35$K) 
hence ferromagnetic and frustrated for both
Ho$_2$Ti$_2$O$_7$ 
and Dy$_2$Ti$_2$O$_7$, but not for
Tb$_2$Ti$_2$O$_7$
for which $J_{\rm nn}\approx -0.9$ K and
$D_{\rm nn}=+0.8$ K
\cite{tb2ti2o7,gingras-tb2ti2o7,tb2ti2o7-jpsj,tb2ti2o7-except}.
  It would therefore appear
natural to ascribe the spin ice behavior in both 
Ho$_2$Ti$_2$O$_7$ and Dy$_2$Ti$_2$O$_7$
to the
positive $J_{\rm eff}$ value as in the simple model of Harris and
Bramwell~\cite{harris-prl,harris-jpc}. However, the situation is much
more complex than it naively appears, especially when considering the
spin ice phenomenon in the context of an underconstrained problem.

The main conceptual difficulty in intuitively understanding the physical
origin of spin ice behavior in rare-earth titanates
stems from the very
nature of
the large dipole-dipole interactions in the Dy$_2$Ti$_2$O$_7$ and
Ho$_2$Ti$_2$O$_7$ materials. Recall the 
general discussion above on the role of perturbations in frustrated magnetic
systems with degenerate ground states.
Dipole-dipole interactions are ``complicated'' in that (i) they are
strongly anisotropic since they couple the spin,
${\bf S}_{i}^{\hat z_{i}}$, and space, ${\bf r}_{ij}$, directions, and
(ii) they are also very long range ($\propto 1/r_{ij}^3$).
One would naively expect that both the spin-space coupling,
(${\bf S}_{i}^{\hat z_{i}}\cdot {\bf r}_{ij}$),
 and the
long range nature of the interaction (beyond nearest neighbor) introduces
``so many'' energetic constraints on the ground state
spin correlations that
consideration of the dipolar interaction beyond nearest neighbor would
lift most, if not all  local degeneracies. Such an effect would  give rise to a
long range ordered state at a well defined (possibly incommensurate)
wave vector at some critical temperature $T_c(J,D) \sim O(D_{\rm  nn}/k_{\rm B})
\sim 2$ K, as opposed to
spin ice behavior.
This simple argument would appear to be further supported by noting that recent
calculations show that vanishingly small, but nonzero long range dipole-dipole
interactions select a unique (non-degenerate) long range N\'eel ordered
state for Heisenberg spins in 
an otherwise classical nearest neighbor pyrochlore
antiferromagnet \cite{gd2ti2o7,palmer}.
This clearly does not happen in  real spin ice materials such as
Ho$_2$Ti$_2$O$_7$~\cite{harris-prl,bramwell-Sq} and
Dy$_2$Ti$_2$O$_7$~\cite{ramirez-nature}, at least 
down to a temperature 
$T\sim 200$ mK.
Hence the question that we address here is:
\\

\noindent{\em
When there is an effective ferromagnetic nearest neighbor
interaction, $J_{\rm eff}$, in $(111)$ Ising pyrochlores, why
do long range dipolar interactions fail to destroy 
spin ice behavior, and not, instead, give rise to long range
N\'eel order with a critical temperature $T_c \sim O(D_{\rm nn})$?
}
\\

A simple explanation to  this question might be that
the inability of long range dipolar interactions to lift the ground state
degeneracy established at the nearest neighbor level
arises from a 
``mutual frustration'' of the degeneracy-lifting energetics, or mean-field, coming
from the
spins
beyond the
nearest neighbor distance. This argumentation, 
if correct, suggests that a mean-field theory which
contains dipolar interactions up to an ``appropriate cut off''
distance,
and aimed at determining the ordering wave vector as a function
of the cut-off distance of the dipolar interactions may be used as a first step
to investigate the microscopic origin of spin ice behavior,
and the failure of
a dipolar-driven 
degeneracy lifting process at a temperature $T\sim O(D_{\rm nn})$.
As we show below, we find that the truncation of dipolar interactions
lead to spurious results when the truncation distance is less than
$\lesssim  100$ nearest neighbors, and that spin ice behavior is restored
only when considering the dipolar interactions to very large cut-off distances.

The rest of the paper is organized as follows. In the next section we present 
a 
description of the mean-field theory we use to determine the ordering
 wavevector
in $(111)$ Ising pyrochlores as a function of cut-off distance. Our results 
are
presented in Section III, followed by a brief discussion in Section IV.

%%%%%%%%%%%%%%%%%%%%%%%%%%%%%%%%%%%%%%%%%%%%%%%%%%%%%%%%%%%%

\section{Mean-Field Theory}
\label{theory}

Our aim in this section is to use
mean-field theory to determine 
the the critical (``soft'') modes and, consequently, the
nature of the magnetic 
phase(s) exhibited by
a classical model of Ising spins on a pyrochlore
lattice with local  axes (the
$(111)$
directions of the cubic unit cell). This system is
described by
a Hamiltonian with
nearest
neighbor exchange and long range dipolar interactions
 \cite{denHertog-prl1,denHertog-loop}:
\begin{eqnarray}
\label{hamiltonian}
H&=&-J\sum_{\langle ij\rangle}{\bf S}_{i}^{{\hat z}_{i}}\cdot{\hat S}_{j}^{
{\hat z}_{j}}
\nonumber \\
&+& Dr_{{\rm nn}}^{3}\sum_{i>j}\frac{{\bf S}_{i}^{{\hat z}_{i}}\cdot{\bf
S}_{j}^{{\hat z}_{j}}}{|{\bf r}_{ij}|^{3}} - \frac{3({\bf S}_{i}^{{\hat
 z}_{i}}\cdot{\bf r}_{i
j})
({\bf S}_{j}^{{\hat z}_{j}}\cdot{\bf r}_{ij})}{|{\bf r}_{ij}|^{5}} \;\; .
\end{eqnarray}
The first term is the near neighbor 
exchange interaction, and the
second term is the dipolar coupling between the $(111)$
Ising magnetic moments.
For the open pyrochlore lattice structure, we expect very small
second and further nearest neighbor exchange
coupling~\cite{greedan-open-pyro}. We therefore
only consider exchange interactions
between nearest neighbor spins.
Here the spin vector ${\bf S}_{i}^{{\hat
z}_{i}}$ labels the Ising moment of
magnitude
 $\vert {\bf S}_{i}^{{\hat z}_{i}} \vert=1$ at lattice site $i$ and
oriented along the
{\it local}
Ising $(111)$
axis ${{\hat z}_{i}}$. The distance $\vert {\bf r}_{ij}\vert$
is measured in
units of the nearest neighbor distance, $r_{\rm nn}$.
Here
$J$ represents the
exchange energy and $D=(\mu_{0}/4\pi)\mu^{2}/r_{\rm nn}^{3}$.
Because of the local Ising axes,
the effective nearest neighbor
energy scales are $J_{\rm nn}\equiv J/3$ and, as
mentioned above, $D_{\rm nn}\equiv 5D/3$, since
${\hat z}_i\cdot{\hat z}_j = -1/3$ and
$({\hat z}_i\cdot{\bf r}_{ij})
 ({\bf r}_{ij}\cdot{\hat z}_j) = -2/3$.

We now proceed along the lines of Reimers, Berlinskly and Shi
in their mean-field study of 
Heisenberg pyrochlore antiferromagnets \cite{reimers-mft}, and
which has recently been extended to investigate systems with
long range dipole-dipole interactions \cite{gd2ti2o7,palmer}.
We consider the mean-field order parameters, ${\bf  B}({\bf  r_i})$ at
site ${\bf r}_i$.
The pyrochlore lattice is a non-Bravais lattice, and we
use a rhombohedral basis where there are four
atoms per unit cell located at
$(0,0,0)$,
$(1/4,1/4,0)$,
$(1/4,0,1/4)$, and
$(0,1/4,1/4)$ 
in units of the conventional cubic unit cell of size $a=r_{\rm nn}\sqrt8$. 
Each of these four points define an FCC sublattice of cubic unit cell 
of size $a$.
We relabel the spins, ${\bf S}({\bf r}_i)$, in terms of unit cell coordinates, 
and a
sublattice index within the unit cell, and
take advantage  of the translational symmetry of the lattice by expanding
the order parameters ${\bf B}(\bf r_i)$ in terms of Fourier
components.
In this case ${\bf  B}^a({\bf  r}_i)=B({\bf r}_i)\hat z^a_i$
 on the $a$'th sublattice site of the
unit cell located at ${\bf r}_i$
can be written as
\begin{equation}
 B^a({\bf r}_i) {\hat z}_i^a  = \sum_{{\bf q}} 
 B^a({\bf q})  {\hat z}_i^a \exp(i{\bf q}\cdot {\bf  r}_i)	\;\;\; ,
\end{equation}
where $\hat z_i^a$ is a unit vector along the local $\langle 111\rangle$
Ising axis on the $i$'th site of sublattice $a$.
The spin-spin interaction matrix, ${\cal J}^{ab}(\vert {\bf 
r}_{ij}\vert)$,
including both exchange and dipolar interactions, reads:
\begin{eqnarray}
{\cal J}^{ab}(\vert{\bf r}_{ij}\vert ) & = & 
J ({\hat z}^a_i\cdot {\hat z}_j^b)
\delta_{r_{ij},r_{\rm nn}}  \\
&+& D_{dd}\left \{
\frac{ {\hat z}^a_i\cdot {\hat z}_j^b } {(r_{ij}^{ab})^3} - 
3
\frac{{\hat z}^a_i \cdot {\bf r}_{ij}^{ab} 
      {\hat z}^b_j \cdot {\bf r}_{ij}^{ab} }
 {(r_{ij}^{ab})^5}
\right \} 
\;,
\end{eqnarray}
where $\delta_{\alpha\beta}$ is the Kronecker delta.
${\bf r}_{ij}^{ab}$ denotes the 
interspin vector ${\bf r}_{ij}$ that 
connects spin ${\bf S}_i^a$ on the $a$ sublattice
to spin ${\bf S}_j^b$ on the $b$ sublattice.
We write ${\cal J}^{ab}(\vert {\bf r}_{ij}\vert$
in terms of its
Fourier components as
\begin{equation}
{\cal J}^{ab}(\vert {\bf r}_{ij}\vert)=\frac{1}{M_{cell}}\sum_{\bf 
q}
{\cal J}^{ab}({\bf q})\exp(-i{\bf q}\cdot {\bf r}_{ij})\;\;\;	.
\label{eq-transform}
\end{equation}
where $M_{cell}$ is the number of unit cells with 4 spins per unit cell.
The quadratic part of the mean-field free-energy, $F^{(2)}$,
then becomes~\cite{reimers-mft}:
\begin{equation}
F^{(2)}(T)/M_{cell}=\frac{1}{2}\sum_{{\bf q},(ab)}
\!\!\! B^a({\bf q}) 
\left\{ T\delta_{ab}-{\cal J}
^{ab}({\bf q})\right\}B^b(-\bf q) 
\;\;\; ,
\end{equation}
where $T$ is the temperature in units of $1/k_{\rm B}$.
Diagonalizing $F^{(2)}(T)$ requires transforming to normal modes of the 
system
\begin{equation}
B^a({\bf q}) = \sum_\alpha
U^{a,\alpha}\Phi^\alpha(\bf q)	\;\;\; ,
\end{equation}
where $\{ \Phi^\alpha(\bf q)\}$ are the eigenmodes, and
$U(\bf q)$ is the unitary matrix that diagonalizes ${\cal J}^{ab}(\bf  q)$
in the sublattice space, with eigenvalues $\lambda^{\alpha}(\bf q)$
\begin{equation}
\sum_b {\cal J}^{ab}({\bf 
q})U^{b\alpha}({\bf  q})=
\lambda^\alpha({\bf q})U^{a\alpha}({\bf  q})
\;\; .
\end{equation}
Henceforth we use
the convention that indices $(ab)$ label sublattices,
and index $\alpha$
labels the normal modes.
We express $F^{(2)}(T)$ in terms of normal modes as
\begin{equation}
F^{(2)}/M_{cell}=\frac{1}{2}\sum_{\bf  q} \sum_\alpha 
\Phi^\alpha({\bf  q})\Phi^\alpha(-{\bf  q})
\left\{ k_{\rm B}T-\lambda^\alpha({\bf q}) \right \}
\;\;\; .
\end{equation} 
An ordered state first occurs at the temperature 
\begin{equation}
T_c = {\rm max}_{{\bf q}} \{\lambda^{max}(\bf q)\}
\;\;\; ,
\end{equation}
where $\lambda^{max}({\bf q})$ is the largest of the four eigenvalues 
($\alpha =1,2,3,4$) at wavevector ${\bf q}$, and
where $\max_{{\bf q}}$ indicates a 
global maximum of the spectrum of $\lambda^{max}({\bf q})$ for all 
${\bf q}$.
The value of ${\bf q}$ for which $\lambda^\alpha({\bf q})$ is maximum is
the ordering wavevector ${\bf q_{\rm ord}}$.

Let us briefly explain how we proceed using the above set of
equations to determine the
critical (``soft'') mode(s) of the system at $T_c$. 
The Fourier transform of ${\cal}J^{ab}(\vert
{\bf r}_{ij}\vert)$ is calculated using the inverse transform of Eq. (2.5).
The pyrochlore lattice has a symmetry of inversion with respect to a
lattice point and this implies that ${\cal J}^{ab}({\bf q})$
is real and symmetric.
The eigenvalues and eigenvectors are found using a standard 
numerical packages for eigen problems of real symmetric matrices.

%%%%%%%%%%%%%%%%%%%%%%%%%%%%%%%%%%%%%%%%%%%%%%%%%%%%%%%%%%%%%%%%%%%%%%%%%%%

\section{Results}

For each ${\bf q}$ there are 4 eigenvalues of ${\cal J}^{ab}({\bf q})$
($\lambda^{\alpha}({\bf q})$, $\alpha=$ 1, 2, 3, 4).  To determine 
${\bf q_{\rm ord}}$
 we need to find the ${\bf q}$ value for
which ${\lambda^{\alpha}}$ is maximum. We therefore calculate
$\lambda^{max}({\bf q})$, the largest of the 4 eigenvalues
of ${\cal J}^{ab}({\bf q})$ for each ${\bf q}$. Figures 2$-$6 show 
$\lambda^{max}({\bf q})$ vs ${\bf q}$ in the $\langle hhl \rangle$ plane
for various cut-off distance, $N_c$, of the dipolar interactions
used in the calculation of ${\cal J}^{ab}({\bf q})$ via the inverse
transform of Eq. \ref{eq-transform}.
Figures 2$-$6 correspond to cut-off distances for the
1st, 2nd, 5th, 100th and 1000th
nearest neighbor ($N_c=1$, 2, 5, 100 and 1000), and which
correspond to physical distances $R_c(N_c=1)/r_{\rm nn} =1$,
				 $R_c(N_c=5)/r_{\rm nn} = \sqrt 7$,
				 $R_c(N_c=10)/r_{\rm nn}= \sqrt 13$,
				 $R_c(N_c=100)/r_{\rm nn}=\sqrt 136$ and
				 $R_c(N_c=1000)/r_{\rm nn}=37$,
 expressed in units of the nearest neighbor distance, $r_{\rm nn}$.
In the calculations here, we have used $J=0$ and measure 
$\lambda^{max}({\bf q})$ in units of the dipolar strength, $D$.
Our conclusions below are independent of the choice made for $J$ as
long as we are in the spin-ice regime $(J/3+5D/3) \gtrsim 0$. For
$J$ sufficiently negative (antiferromagnetic) the ordering wavevector
is at ${\bf q_{\rm ord}}=0$,
 corresponding to a two-fold Ising state all-in$-$all-out
as discussed in 
Refs.\cite{bramginreim,harris-jpc,moessner-rapcomm,denHertog-prl1}.
Several comments are in order.

Cutting off the dipolar interaction at $N_c=1$ gives the effective
nearest neighbor Hamiltonian:
\begin{equation}
H_{\rm eff}(N_c=1) \;=\; \left ( \frac{J}{3}+\frac{5D}{3} \right )
		\sum_{\langle i,j\rangle} \sigma_i \sigma_j 
	\;\;\; ,
\end{equation}
where $\sigma_i=\pm 1$.
For $J_{\rm eff}=(J+5D)/3=J_{\rm nn}+D_{\rm nn}>0$, and
 we recover the nearest neighbor
Ising spin ice model of Harris and Bramwell \cite{harris-prl,harris-jpc}.
As discused in Refs. \cite{harris-prl,harris-jpc}, this model maps onto
Anderson's antiferromagnetic Ising model with global $\hat z$ quantization
axis \cite{anderson}.
 Consequently,
it is normal to recover in Fig. 2 the spectrum of $\lambda^{max}({\bf q})$
found for the pyrochlore antiferromagnet \cite{reimers-mft}.
Indeed, the largest eigenvalue found in Ref. \cite{reimers-mft} is
$2 J_{\rm eff}$. Using $J_{\rm eff}=J_{\rm nn}+D_{\rm nn} = 5D/3$, we get the
flat spectrum at $\lambda({\bf q})/D = 10/3$, as in Fig. 2.
(In Figs. 2-7, ${\bf q}$ is measured in units of $2\pi/a$).
The flatness of $\lambda^{max}({\bf q})$ reflects the ${\bf q}-$space
representation of the zero energy excitation modes in  real
space. In the kagom\'e Heisenberg antiferromagnet lattice, these are the
so-called weathervane modes~\cite{chandra,shender}. Here, 
for Ising spins, they correspond to closed loops of equal in and out
spins which obey the ice-rules\cite{anderson,bernal,pauling}. These 
were first identified by
Anderson \cite{anderson}, and
have recently been exploited in computer simulations of the dipolar
spin ice model~\cite{denHertog-loop}.

In Fig. 3 we see that $\lambda^{max}({\bf q})$ develops an absolute
maximum for $N_c=5$.
This means that a well defined mode becomes soft (massless)
at the corresponding incommensurate
${\bf q_{\rm ord}}$,
 which should give rise to incommensurate 
long range N\'eel order, unlike what was 
found in Refs. \cite{ramirez-prl,siddharthan}
for their choice $N_c=5$. We comment on this further in Section IV.
 
For $N_c=10$ (Fig. 4), we see that ${\bf q_{\rm ord}}$
 is different from
its value for $N_c=5$. Clearly, ${\bf q_{\rm ord}}$
 is not a rapidly
converging function of $N_c$.  We note also that
some of the flatness present for $N_c=1$ shown in Fig. 2 is being
recovered 
for $N_c=10$.
For $N_c=100$ (Fig. 5), we clearly
note modulations (``ripples'') in
$\lambda^{max}({\bf q})$ in the $\langle 111\rangle$ direction, with now
increased
flatness restored in the overall spectrum.
As $N_c$ is increased, these modulations ``interfere'' in such a manner
as to restore a very smooth surface, as shown in Fig. 6 for $N_c=1000$.
This is similar to what is found in the classical Heisenberg pyrochlore
lattice with long range dipole-dipole interactions where the 
amplitude of the ripples in
the $(111)$ directions are continuously decreasing and mutually cancelling out
each other as the number of nearest neighbor distances considered in the dipolar
lattice sum is increased to infinity and where, in the limit $N_c\rightarrow\infty$,
a degeneracy line occurs at the star of $(111)$ 
\cite{gd2ti2o7,palmer}.
Essentially, the ripples get smoothed out and pushed to zero as the 
finite-size effect of the  cut-off wavevector $q_c \sim 2\pi/R_c(N_c)$ goes to zero.
This is more explicitely shown in Fig. 7 where
 $\lambda^{max}(hh0)$ is shown.
The convergence at strictly ${\bf q}=0$
is slow since the dipolar
lattice sum is
conditionally convergent for this ${\bf q}$ value.
This is the origin of the downward ``spike'' in Fig. 7 for
${\bf q}=0$.

%\newpage

\begin{figure} \begin{center}
\includegraphics[width=7cm]{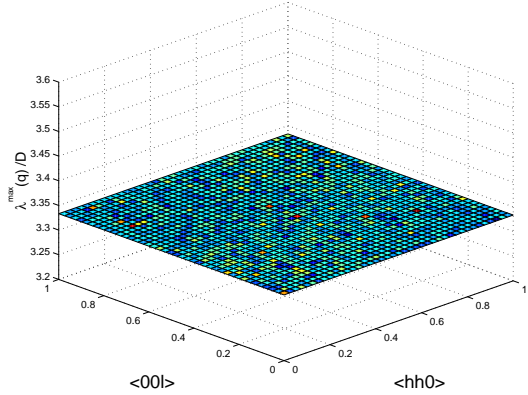}
\vspace{7mm} 
\caption{$\lambda^{max}({\bf q})$ vs
${\bf q}$ in the $\langle hhl \rangle$ plane for
$N_c=1$.} 
\end{center} \end{figure}

\begin{figure} \begin{center}
\includegraphics[width=7cm]{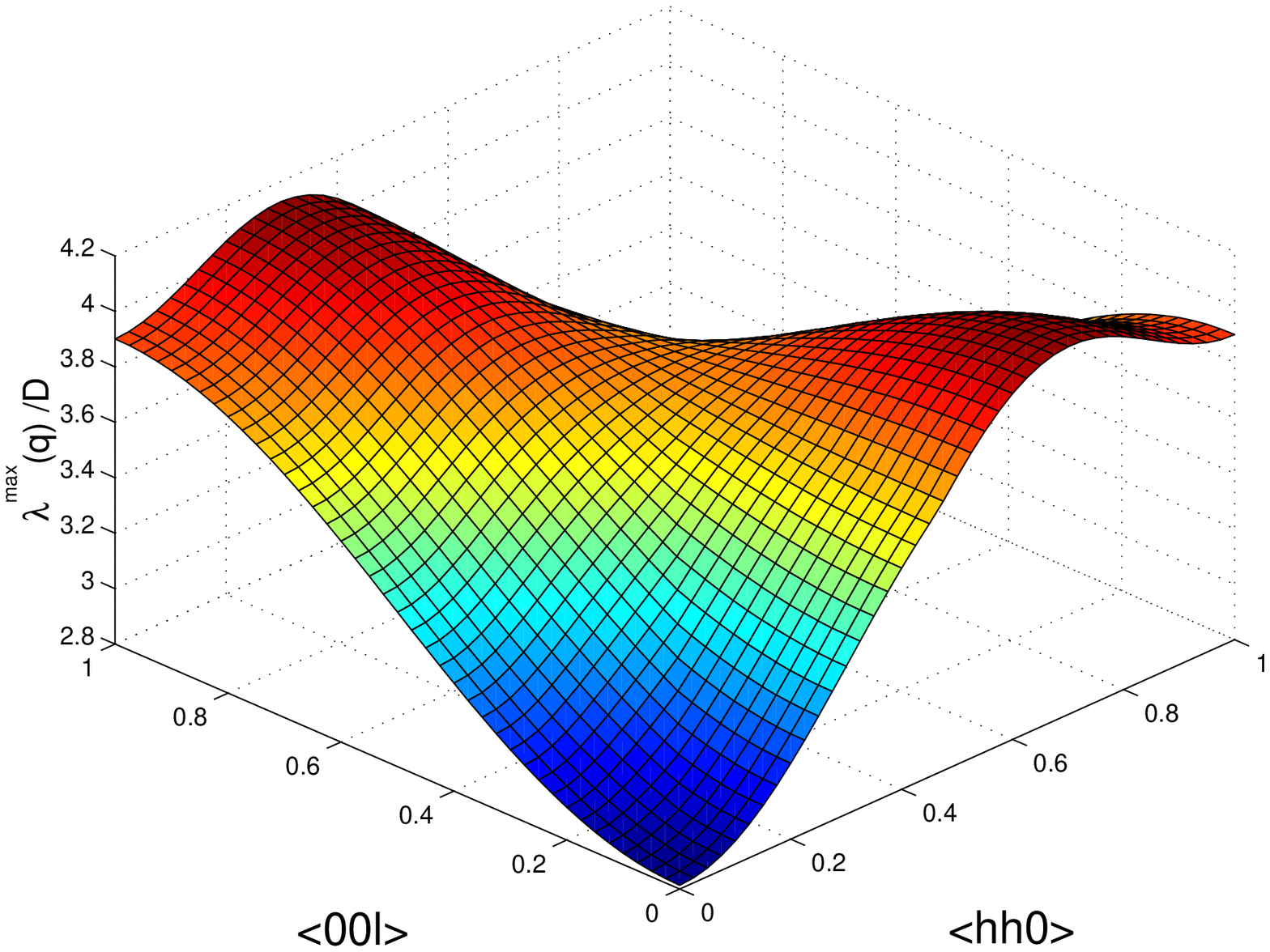}
\vspace{7mm} 
\caption{$\lambda^{max}({\bf q})$ vs
${\bf q}$ in the $\langle hhl \rangle$ plane for
$N_c=5$.} \end{center} \end{figure}

\begin{figure} \begin{center}
\includegraphics[width=7cm]{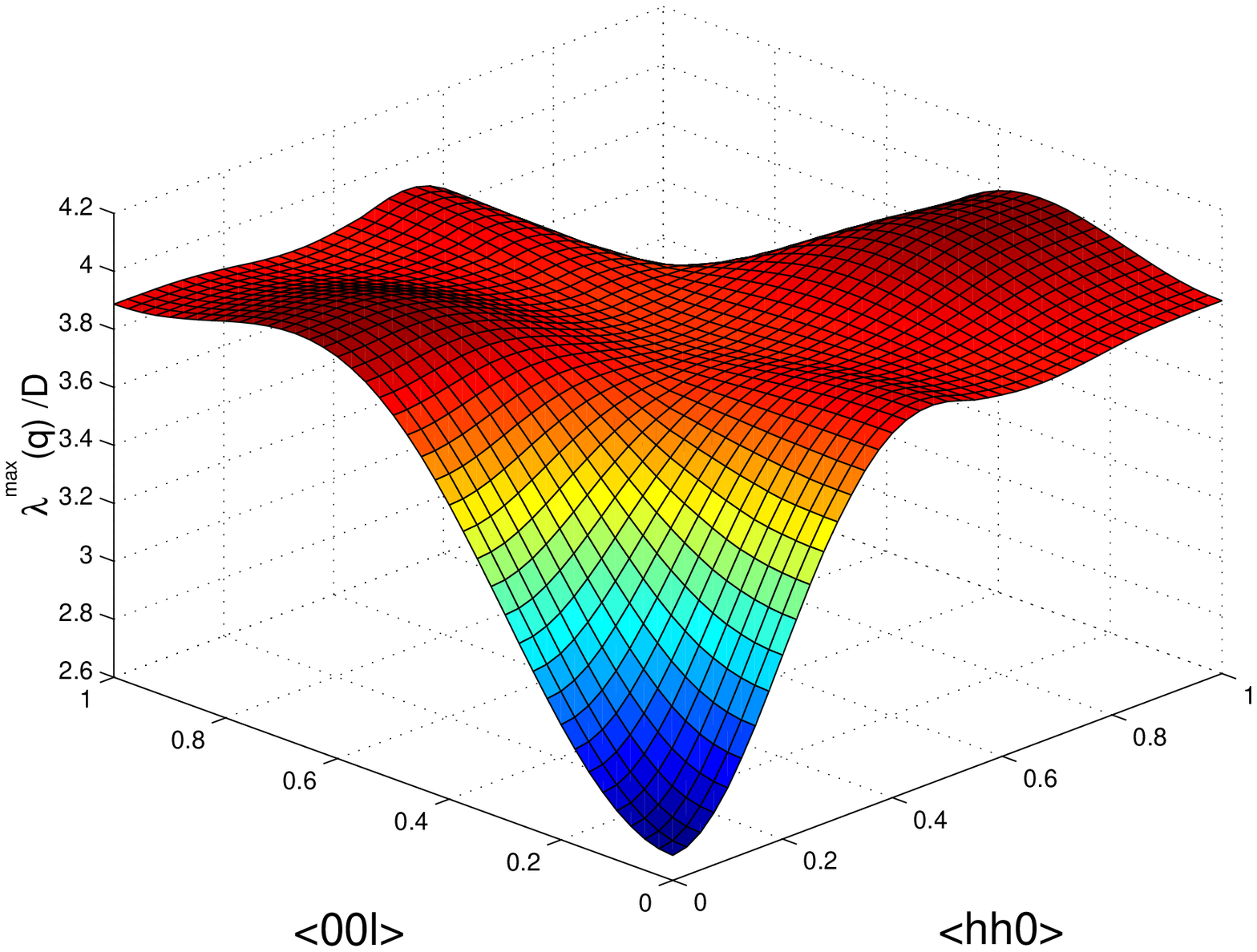}
\vspace{7mm} 
\caption{$\lambda^{max}({\bf q})$ vs
${\bf q}$ in the $\langle hhl \rangle$ plane for
$N_c=10$.} \end{center} \end{figure}

\begin{figure} \begin{center}
\includegraphics[width=7cm]{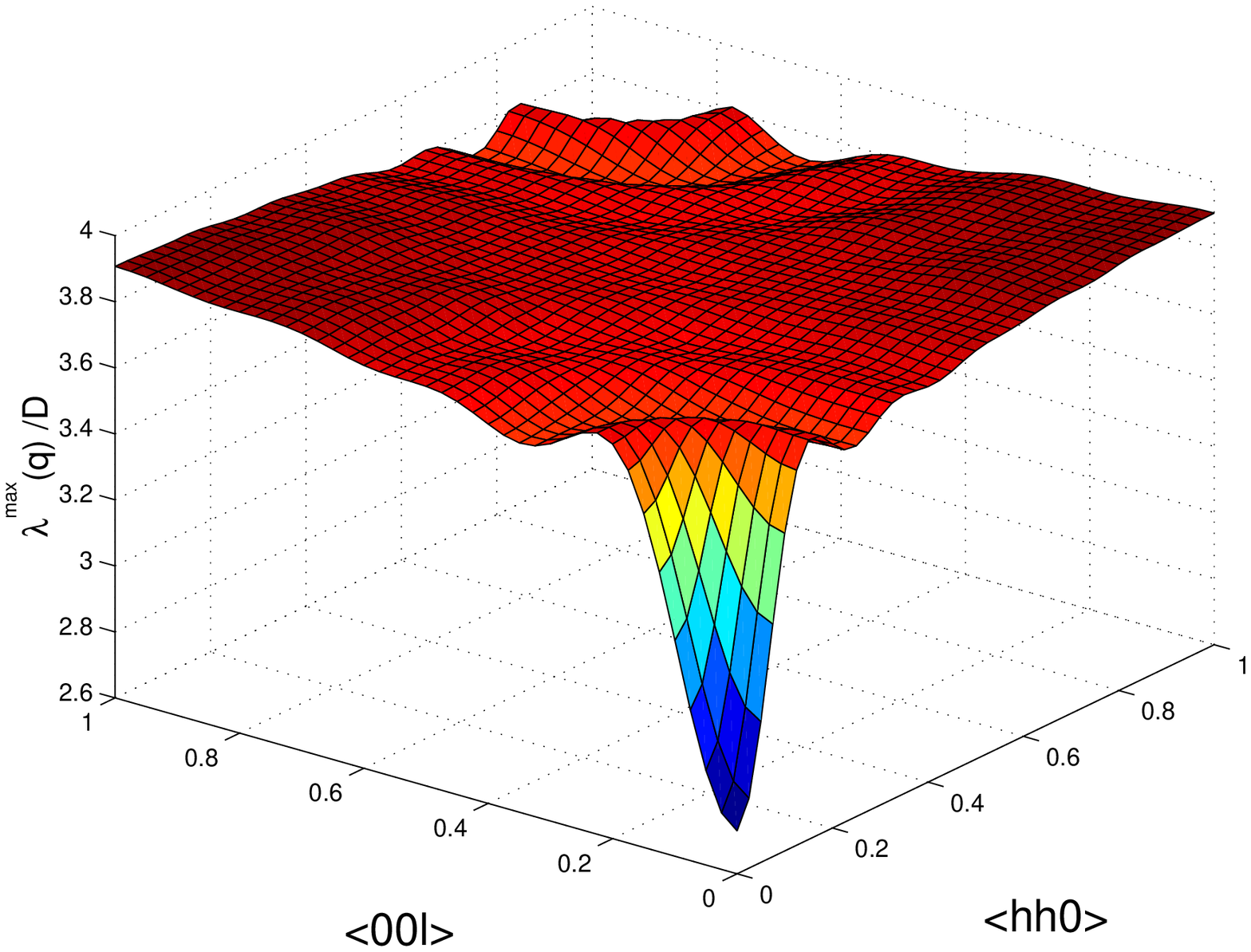}
\vspace{7mm} \caption{$\lambda^{max}({\bf q})$ vs
${\bf q}$ in the $\langle hhl \rangle$ plane for
$N_c=100$.} \end{center} \end{figure}

\begin{figure} \begin{center}
\includegraphics[width=7cm]{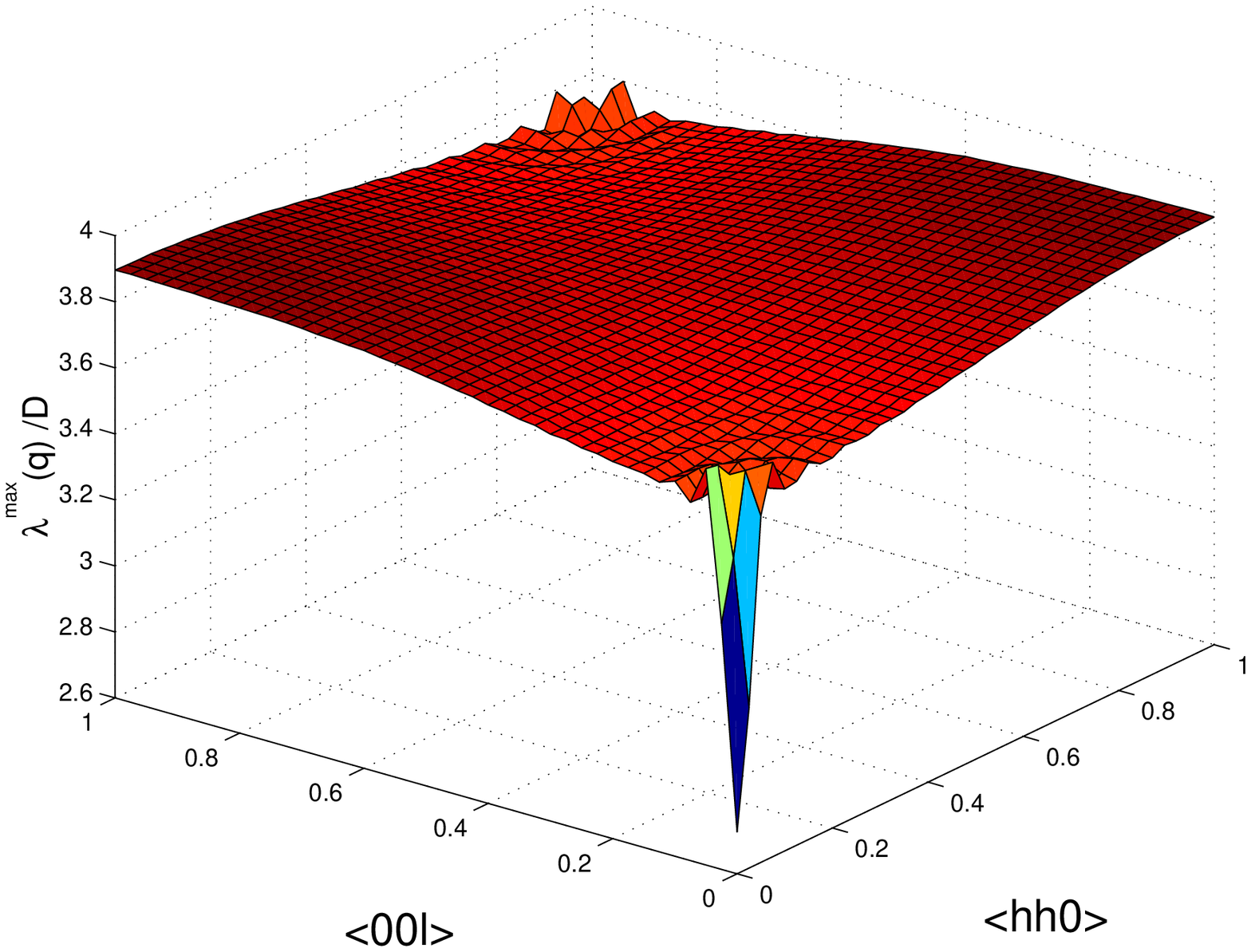}
\vspace{7mm} 
\caption{$\lambda^{max}({\bf q})$ vs
${\bf q}$ in the $\langle hhl \rangle$ plane for
$N_c=1000$.} \end{center} \end{figure}

\begin{figure} \begin{center}
\includegraphics[width=7cm]{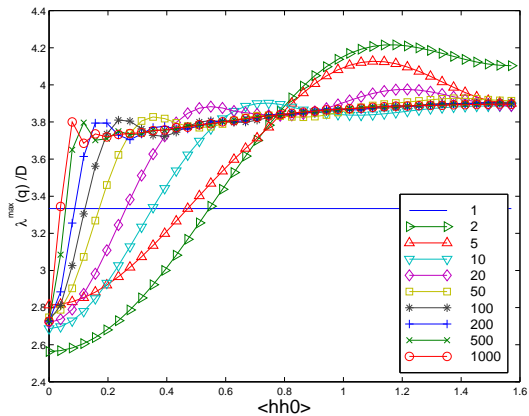}
\vspace{7mm} 
\caption{$\lambda^{max}({\bf q})$ vs
${\bf q}$ for ${\bf q}$ in the $\langle 100\rangle$ direction, 
for 
$N_c=1$ (solid line, no symbols), 
$N_c=2$ (right triangles),
$N_c=5$ (up triangles), 
$N_c=10$ (down triangles),
$N_c=20$ (diamond), 
$N_c=50$ (squares), 
$N_c=100$ (stars), 
$N_c=200$ (pluses), 
$N_c=500$ (crosses), and
$N_c=1000$ (circles).
} \end{center} \end{figure}

\section{Discussion}

Figure 6 constitutes our main result. We see that as $N_c\rightarrow \infty$,
the spectrum of $\lambda^{max}({\bf q})$  becomes very flat,  similar but
not identical to the nearest neighbor spin ice model (Fig. 2).
This shows that it is the very long distance behavior of dipolar interactions
leads to a high frustration that
 restores the symmetry and loops as low-energy 
excitations obeying the
ice-rules \cite{anderson,denHertog-loop},
 and, therefore, causes the spin ice phenomenon in Ising pyrochlore
rare-earths systems.
One notes that  $\lambda^{max}({\bf q})$   displays a weak maximum
at ${\bf q_{\rm ord}}
= (2\pi/a)[1,0,0]$.
The low-lying maximum and weak dispersion of $\lambda^{max}({\bf q})$
when $N_c\rightarrow \infty$ with respect to the overall smooth and
flat profile of the spectrum implies a small ordering temperature,
$T_c$, compared to the nearest neighbor dipolar energy scale, $D$,
in agreement with our recent loop Monte Carlo simulations where
we found $T_c/D \sim 0.11$ \cite{denHertog-loop}. Also, ${\bf q_{\rm ord}}$
found here in the mean-field theory for $N_c \rightarrow \infty$
is the same ordering
 wave vector as the one  found in our Monte Carlo simulations
\cite{denHertog-loop}.
Consequently, the answer to the question raised in the
Introduction does appear to be that there is indeed a
high level of frustration in spin ice materials that come
from the
energetics beyond nearest-neighbor. However, this frustration
is not perfect, and there is indeed a lifting of the degeneracy
caused by the dipoles, but only at a temperature scale
$T/D \sim 0.1$.

The results presented here of the quasi-degeneracy
in $\lambda^{max}({\bf q})$, and the consequential spin ice
behavior being recovered when $N_c \rightarrow \infty$ (as
found
in our Monte Carlo simulations \cite{denHertog-prl1,denHertog-loop} also),
raise some questions as to the validity of the simulation work
of Refs. \cite{ramirez-prl} and \cite{siddharthan}
   where 
dipole-dipole interactions were cut-off for $N_c=5$ \cite{ramirez-prl}
and $N_c=12$ \cite{siddharthan}. 
We believe that the results
of Refs. \cite{ramirez-prl} and \cite{siddharthan} are invalid as pertaining
to real materials due to
the truncation used for the dipolar sum.
Specifically, for the values used in Ref. \cite{ramirez-prl} to 
model Ho$_2$Ti$_2$O$_7$, and with a choice of $N_c=5$, we have found
in Monte Carlo simulations
that the paramagnetic scattering is indeed incommensurate \cite{parascat}, in
agreement with our mean-field calculations. The
constraint to work with a small 
number of unit cells 
and system size $L$ incompatible
 with an incommensurate
 ${\bf q_{\rm ord}}$ that would be selected for $N_c=5$ for a
thermodynamically large sample
 may precipitate
the system into a partially ordered state 
via some type of freezing transition as found in
Ref.\cite{ramirez-prl}.
Conceptually,
this is similar to the difficulties in finding the
appropriate long range ordered vortex lattice ground state
in Monte Carlo simulations of
dense frustrated Josephson junction arrays 
 in a magnetic
field, and where 
glassy behavior is found when too small system
sizes are considered \cite{gupta,denniston}.
At any rate, what happens in a simulation with $N_c \lesssim O(10^2)$
nearest neighbors is a moot issue in terms of modeling and understanding
spin ice in real materials as our analysis here shows.

In conclusion, we have shown that one can understand at the mean-field
level that it is the true long range nature of dipolar interactions
that cause the quasi-degenerate ice-rules obeying states in 
Ising pyrochlore magnets such as Ho$_2$Ti$_2$O$_7$ and 
Dy$_2$Ti$_2$O$_7$. However, similar to the so-called energetic
ice models~\cite{lieb,energy-ice}, dipolar interactions do slightly energetically
select a preferred state with long range N\'eel order at 
very low temperature.  It would be interesting and useful to understand
what is the specific
relationship between the symmetry of long range dipolar interactions
for local $(111)$ Ising spins coupled via long range dipolar interactions,
 and the symmetry of the
pyrochlore lattice that leads to an asymptotic restoration of
a approximately ice-rules obeying 
quasi-degenerate ground state manifold.

\acknowledgements

\noindent We thank Steve Bramwell and Peter
 Holdsworth for stimulating discussions.  
We acknowledge Maxime Dion and Roger Melko for collaborations on these
and related studies.
This reserach has been funded by NSERC of Canada.  M.G. acknowledges the
Research Corporation for a Research Innovation Award and a Cottrell
Scholar Award, and the Province of Ontario for a Premier Research
Excellence Award.

%\newpage


\begin{thebibliography}{99}
\bibitem{tb2ti2o7-except}Tb$_2$Ti$_2$O$_7$ is a more complicated case
than Dy$_2$Ti$_2$O$_7$ and Ho$_2$Ti$_2$O$_7$.  This system is
not a perfect $(111)$ Ising model since $\theta_{\rm CW} \sim -20$ 
is not small compared to the low-lying single ion crystal electric
field levels at energies of $\sim 20$ K and $\sim 100$ K
above the ground state Ising doublet \cite{tb2ti2o7,gingras-tb2ti2o7}.
\end{thebibliography}
\end{document}